%% file: network_formation.tex
\documentclass[12pt]{article}

\usepackage{verbatim}
\usepackage{natbib}
\usepackage{amsfonts}
\usepackage{amssymb}
\usepackage{amsmath}
\usepackage{amsthm}
\usepackage{mathrsfs}
\usepackage{algorithm}
\usepackage{algorithmic}
\usepackage{enumerate}
\usepackage{hyperref}
\usepackage{xr-hyper}
\usepackage[pdftex]{color,graphicx}
\usepackage[margin=1.3in]{geometry}
\usepackage{setspace}
\usepackage{adjustbox}
\usepackage{booktabs}

\newtheorem{proposition}{Proposition}
\newtheorem{lemma}{Lemma}
\newtheorem{ass}{Assumption}

\newcommand{\bs}{\boldsymbol}
\def\bal#1\eal{\begin{align}#1\end{align}}
\def\bals#1\eals{\begin{align*}#1\end{align*}}
\def\be#1\ee{\begin{equation}#1\end{equation}}

\DeclareMathOperator*{\argmin}{argmin\;}

\title{Causal inference for social network formation\thanks{
We thank Peng Ding, Bryan Graham, and Aureo de Paula for valuable feedback and suggestions.\vspace{12pt}
}}
\author{Maximilian Kasy\footnote{Department of Economics, University of Oxford. \href{mailto:maximilian.kasy@economics.ox.ac.uk}{maximilian.kasy@economics.ox.ac.uk}.} \and Elizabeth Linos\footnote{Harvard Kennedy School. \href{mailto:elizabeth_linos@hks.harvard.edu}{elizabeth\_linos@hks.harvard.edu}.} \and Sanaz Mobasseri\footnote{School of Management, University College London.\href{mailto:s.mobasseri@ucl.ac.uk}{s.mobasseri@ucl.ac.uk}.}}

\begin{document}
\maketitle
\onehalfspacing

\begin{abstract}
\input{Sections/0_abstract}\\[12pt]

\end{abstract}

\clearpage
\input{Sections/1_introduction}

\input{Sections/2_setup}

\input{Sections/3_estimation}

\input{Sections/4_background}

\clearpage
\input{Sections/5_empirical}

\clearpage
\bibliographystyle{econsoc}
\bibliography{library}

\end{document}

%% file: Sections/0_abstract.tex
This paper develops a framework for identification, estimation, and inference on the causal mechanisms driving endogenous social network formation.
Identification is challenging because of unobserved confounders and reverse causality; inference is complicated by questions of equilibrium and sampling. 
We leverage repeated observations of a network over time and random variation in initial ties to address challenges to causal identification. 
Our design-based approach sidesteps questions of sampling and asymptotics by treating both the set of nodes (individuals) and potential outcomes as non-random.
We apply our approach to data from a large professional services firm, where new hires are randomly assigned to project teams within offices.
We estimate the causal effect on tie formation of indirect ties, network degree, and local network density. 
Indirect ties have a strong and significant positive effect on tie formation, while the effects of degree and density are smaller and less robust.

%% file: Sections/1_introduction.tex
\section{Introduction}
\label{sec:introduction}

Social networks play a central role in shaping economic and social outcomes for both their constituent members and for society at large. 
Networks determine the diffusion of information and misinformation, provide access to jobs and opportunities, and mediate the spread of financial shocks and infectious diseases. 
A large empirical literature has documented that network features---such as density, centrality, and size---are highly predictive of individual behavior and aggregate outcomes.
Yet we know less about the causal mechanisms driving the formation of social ties, and thus networks. 

This paper develops a design-based framework for identification, estimation, and inference on causal mechanisms of social network formation. 
Our approach requires random variation of initial network structure and panel observations of subsequent network evolution. 
Together, these allow us to identify how initial network structure causally affects the probability that future ties form, without imposing an equilibrium model of network formation or restrictions on treatment effect heterogeneity.
We illustrate our approach using data from a large professional services firm, where new employees are randomly assigned to initial project teams within offices, generating quasi-experimental variation in their early network positions.

We study three endogenous network characteristics predicted by theory to affect tie formation: initial exposure to indirect connections, degree, and local network density. 
Our estimands measure how these initial network characteristics causally affect the probability that a future tie forms between two individuals.

\paragraph{Methodological challenges}
Disentangling the causal mechanisms impacting tie formation is notoriously challenging, for at least four reasons \citep{graham2020econometric, de2020econometric}. The first challenge is unobserved confounders. Suppose, for example, we observe that friends-of-friends are often connected themselves. This could be due to some shared unobserved characteristics; people with similar cultural tastes might preferentially form ties, for instance. But it could also be due to triadic closure (i.e., a causal effect of indirect ties).

The second challenge is reverse causality (or simultaneity). Continuing with the above example, a friendship triangle between A, B, and C could exist because A introduced B and C, B introduced A and C, or C introduced A and B. Separating which ties impacted which is necessary for understanding the causal mechanism. 

The third challenge is how to define an appropriate notion of equilibrium for the formation of networks. Nash equilibrium is typically not satisfactory in the network context because it only allows for individual deviations. However, tie formation involves joint decisions among multiple individuals. 
But it is not clear to what extent individuals can coordinate the formation or dissolution of ties. Even for more restrictive notions of equilibrium, such as pairwise stability, equilibrium multiplicity is pervasive and equilibrium selection ambiguous.

The fourth challenge is sampling and statistical inference. Do we consider the observed social network to be randomly sampled from a population of networks, such that we effectively only have one observation? Or do we consider the observed network to be a subnetwork of a larger network, which raises the question of how individuals were sampled from this larger network? 
Both approaches require the researcher to posit a superpopulation of nodes or networks as well as a sampling mechanism, neither of which has a clear empirical counterpart.

\paragraph{Our approach to identification and inference}
In this paper, we propose an approach that addresses these challenges to identification and sidesteps questions of equilibrium and sampling. Our approach has two key requirements. First, we need to observe the network for at least two time periods. 
This allows us to study the dynamics of tie formation, rather than relying on the assumptions of a static equilibrium model.
Second, we require that the initial network is a random draw from a set of networks that can be obtained by a permutation of some individuals.\footnote{We draw inspiration from \citet{randomizationpeer2024} who develop a design-based framework that leverages random permutations of group assignments to identify peer effects and construct exact finite-sample permutation tests---a methodological contribution foundational for our own approach. The key distinction is in the estimand: \citeauthor{randomizationpeer2024} study \emph{peer effects}---the causal influence of peers on individual outcomes---whereas we study causal effects on \emph{network formation itself}, asking how initial network structure shapes the probability that future ties form.} In our empirical application, new hires are randomly assigned to project teams within offices. As a consequence, each permutation of new hires within offices is equally probable.
Such random assignment generates exogenous variation for causal identification.

In addition to these two features of the empirical setting, identification requires exclusion restrictions. We need to make assumptions about which initial network properties matter for the subsequent formation of a tie between a given pair of individuals. We might, for example, assume that formation of such a tie can depend on whether the two individuals concerned initially had any friends in common, but not on any other endogenous network features.

Under these conditions, we derive unbiased estimators for SATEs, finite-sample exact hypothesis tests (for the sharp null of no effects), and conservative confidence intervals (for SATEs). We consider two alternative approaches for estimation. Our first approach is based on inverse probability weighting. Using the (known) permutation structure, we can calculate the probability distribution of the treatment (i.e., the relevant initial network properties) for each tie. Weighting observations using these probabilities makes treatment independent of potential outcomes. Weighted regressions of tie formation on treatment therefore yield unbiased estimators of SATEs.
A shortcoming of this inverse probability weighting approach is that it relies on strong support restrictions: Treatment effects are only identified for pairs of individuals for whom any value of treatment could be obtained under some feasible permutation of the initial network. Weakening this requirement, we consider a second approach based on within-regressions. We show that this second approach identifies a local average treatment effect (LATE).

For either approach, we can calculate p-values for the null hypothesis of no treatment effects, using the permutation distribution of initial network structure. These p-values condition on the sample of individuals, and therefore rely neither on any assumptions about sampling (of individuals from a superset of individuals, or of networks from a superset of networks), nor on asymptotic approximations.
Additionally, we build on results from the literature \citep{tian2025stratified} to show that our estimators are approximately normally distributed over the permutation distribution (conditioning on individuals and potential outcomes).\footnote{The stratified permutational Berry–Esseen bounds proven by \cite{tian2025stratified} provide an explicit bound on deviations from normality, where these deviations becomes negligible both for the case of many offices, and for large offices, in our setting. We thank Peng Ding for pointing us to these results.}

Further, we can estimate upper bounds on standard errors over the permutation distribution, specializing the approach of \cite{mukerjee2018using}, and using the variance of estimates across offices. These bounds are sharp absent effect heterogeneity across offices. In general, only bounds are identified, because standard errors depend on the joint distribution of potential outcomes across arms, as already recognized by \cite{neyman1923application}. Using approximate normality and bounds on standard errors, we construct conservative confidence intervals.

Lastly, we show how our estimators and tests can be calculated efficiently, even in large social networks, by appropriately sequencing calculations and avoiding repeated calculations.

Our approach has a number of advantages.
First, our approach is design-based \citep[cf.][]{abadie2020sampling, abadie2023should}: we condition on the given sample of individuals and potential outcomes, allowing us to construct unbiased estimators and finite-sample valid p-values without relying on asymptotics or hypothetical super-populations.\footnote{Conditioning on potential outcomes, as we do, implicitly also conditions on possible equilibrium selection mechanisms. This raises some subtle issues that will be discussed at the end of \autoref{sec:setup}. We thank Aureo dePaula for pointing this out.}
Second, by leveraging network dynamics rather than a static snapshot, we sidestep questions of equilibrium without having to assume any particular equilibrium notion for tie formation.
Third, our framework allows for unrestricted heterogeneity of causal effects: The effect of local network structure on whether any given pair of individuals forms a tie can vary arbitrarily across pairs. This stands in contrast to parametric or semi-parametric models of network formation, which necessarily restrict heterogeneity, and which often need to assume away sources of endogeneity. 
Our estimands are network analogs of SATEs: Like the SATE in conventional RCTs, they condition on the realized sample of individuals and potential outcomes.

\paragraph{Empirical application}
We apply the proposed approach to data on the intraorganizational employee network of a global professional services firm, ConsultCo.
Building on seven years of administrative employment and staffing data, we construct network ties based on which employees work on the same project at the same time.
When new hires join ConsultCo, they are randomly assigned to their first projects within their office by a centralized human resource (HR) process. 
After this initial placement, tie formation shifts to an informal process that involves a bilateral choice between junior new hires and more senior employees: Junior employees seek out attractive projects, and senior colleagues decide whom to work with. This informal process makes later tie formation endogenous.
Our main estimation sample contains 6,042 employees who were newly hired between 2015 and 2019, and over 130 million pairs of newly hired and more senior employees.

We consider three endogenous network characteristics that might impact the formation of future ties: Indirect ties, node degree, and local network density.
To validate our identifying assumptions, we run a series of placebo tests, based on whether any of these network characteristics appear to impact pre-determined demographic characteristics, including gender, race, and graduation from a top-20 institution. 
We find no effect on any of these pre-determined characteristics, which corroborates our identifying assumptions and in particular the random assignment of new hires to projects within offices.

We next consider the effect of each of indirect ties, node degree, and local network density separately, in binarized form.
Each of those has a strong and highly significant effect when considered in isolation. 
The presence of an indirect tie increases the probability of tie formation by about 40\%, while new hires with above-median initial degree are 23\% more likely to form ties, and above-median local density decreases tie formation by about 18\%.
If we consider a joint model allowing for all three endogenous characteristics to impact tie formation, we find that the effect of indirect ties is quite robust, while the effect of the other two characteristics is reduced.
Similarly, when considering continuous (rather than binary) versions of these endogenous characteristics, only indirect ties appear to have an effect.

\paragraph{Literature}
Decades of research across the social sciences have converged on a remarkably small set of mechanisms that might determine tie formation.
First, opportunity structures---created by institutional settings, organizational contexts, shared foci, and repeated interactions---shape when, where, and with whom people can feasibly form ties~\citep{Blau1977-vb, Feld1981-lc, Sacerdote2001-kp}. 
Second, people are more likely to form ties with others who are similar to them---whether in background (e.g., demographic characteristics), attitudes, or behavior---because similarity eases coordination, increases trust, and reflects people's preferences ~\citep{McPherson2001-pw, currarini2009economic}.
Third, people may form ties for instrumental reasons, weighing the costs and benefits of connection and seeking access to information, resources, or influence~\citep{bala2000noncooperative, Lin2002-hw, Burt2004-uh, jackson2008social, goyal2023networks}.
Last, cognitive and perceptual processes shape tie formation by constraining who people notice, remember, or approach: individuals tend to favor familiar and cognitively ``easy'' categories ~\citep{Dunbar1992-oz, Krackhardt1987-po, Freeman1992-ot, Smith2020-wo}.
Empirical work across diverse settings---from schools ~\citep{Kossinets2006-je, Dahlander2013-ae} to firms ~\citep{Kleinbaum2013-tv} to online platforms \citep{ugander2011anatomy, Lewisetal2012, mosleh2021shared}---consistently corroborates these mechanisms.

Beyond these contextual and individual forces, endogenous structural characteristics of the network are also known to drive how ties form.  
First, theories of triadic closure and transitivity suggest that two people are more likely to form a connection when they share mutual contacts ~\citep{cartwright1956structural, Simmel1964-nr, Heider2013-gy}.
Shared contacts not only create more opportunities for interaction but also reduce the perceived risks of engaging with someone new, as mutual contacts can provide a basis for trust and accountability ~\citep{Coleman1988-ok}.
Second, people with more connections or greater visibility are more likely to attract new ties, creating a tendency for small advantages and disadvantages to compound over time ~\citep{Merton1968-pc, Barabasi1999-hj, DiPrete2006-ua, DiMaggio2012-th}.
Conversely, highly dense networks can sometimes inhibit new tie formation, as existing relationships absorb time and attention and limit exposure to new contacts ~\citep{Burt2009-oh, Uzzi1997-wu, Gargiulo2000-kg}.
Although these endogenous network mechanisms are well developed in theory and often recognized as playing a role in empirical studies of network evolution, there is relatively little causal evidence on their magnitude (for a recent notable exception see \citealt{mosleh2025tendencies}, who provide causal estimates of triadic closure using an online experiment). 

Motivated in part by these sociological contributions, a large literature in mathematics and statistics has developed models of network formation \citep{kolaczyk2014statistical}. One such class of models is exponential random graph models (ERGMs); effectively multinomial logit models at the level of the network, based on arbitrary network characteristics. 
Another class is stochastic block models. 
Both of these are quite general as {descriptive} sampling models of network formation, so that they can generate almost arbitrary networks. But both classes of models are descriptive rather than causal.

In structural econometrics, there is a literature estimating models of network formation assuming individually optimal tie formation and some notion of equilibrium (such as pairwise stability) \citep{graham2015methods, de2018identifying, graham2020econometric, de2020econometric}, based on a static snapshot of a network. 
Such models are causal but often rely on strong assumptions, such as the absence of unobserved confounders. 
A related class of models is based on (myopic) dynamics of tie formation; such models do not require assumptions about equilibrium and equilibrium selection \citep{christakis2020empirical}.

As noted above, we use a design-based approach using random initial variation and network dynamics. In doing so, we build on a long tradition of design-based inference in randomized experiments \citep{neyman1923application}; see \cite{abadie2020sampling, abadie2023should} for more recent expositions.
One notable example using such a design-based approach is \cite{randomizationpeer2024}, which uses random permutations of group assignments to identify (exogenous) peer effects, and to construct exact hypothesis tests.

\paragraph{Roadmap}
The rest of this paper proceeds as follows.
\autoref{sec:setup} introduces our formal assumptions, discusses three network characteristics, and relates our setup to the structural econometrics literature.
\autoref{sec:estimation} then proves identification, derives alternative estimators, shows validity of randomization inference, approximate normality, and standard errors, and discusses computational implementation.
\autoref{sec:background} provides background on our empirical application and data construction.
\autoref{sec:empirical} discusses our empirical findings.
\autoref{sec:proofs} contains all proofs, \autoref{sec:additionalplots} shows additional plots for our empirical application, and \autoref{sec:additionaldiscussion} connects our framework to conventional RCTs.

%% file: Sections/2_setup.tex
\section{Setup}
\label{sec:setup}

In this section we first describe our general setup, which is summarized in \autoref{ass:networkmodel}, \autoref{ass:permutation}, and \autoref{ass:rectangular}. 
We then discuss examples of endogenous tie formation mechanisms, including triadic closure, cumulative advantage, and the effect of local network density.

\paragraph{Adjacency matrices and causal relationship}
Social networks are represented by {adjacency matrices} $A^t$, where $A^t_{ij} \in \{0,1\}$ indicates the presence of a tie between $i,j\in\{1,\ldots,n\}$ at time $t$. 
We are interested in the evolution of social networks, that is, the formation of ties.
The formation of ties might be causally impacted by both observed and unobserved exogenous factors, but also by the presence of pre-existing ties, which are endogenous. 
Because of this endogeneity of tie formation, we have to address questions of reverse causality and the reflection problem \citep{manski1993identification}.

We model the evolution of $A^t$ over time, where we consider networks $A^t$ for times $t=1,2$.
We assume that this evolution is governed by the following {causal relationship}:
\be
  A^2 = f(A^1).
  \label{eq:causal}
\ee
Any sources of either observed or {unobserved heterogeneity}, and in particular any external factors that impact tie formation, are subsumed in the definition of $f$. Any probability statements in the following will condition on $f$, and thus condition on unobserved heterogeneity; all randomness derives solely from the draw of $A^1$ from some set $\mathcal{A}$. 
The relationship in \autoref{eq:causal} is causal because we assume that the function $f$ remains invariant under interventions that change $A^1$.
Our goal is to learn some properties of the mapping $f$.

\paragraph{Repeated observation and exogenous variation}
We observe one realization of both $A^1$ and $A^2$, that is, we have repeated observations of the network over time.
We furthermore assume that there is some {exogenous randomness} in $A^1$: The matrix (network) $A^1$ is sampled uniformly at random from a known set of adjacency matrices $\mathcal A$, {conditional on $f$}.

In our empirical application, new hires are randomly assigned to projects within offices, so that $\mathcal A$ contains all networks obtained by {permutations} of subsets of individuals. Let $\pi$ be a permutation of $\{1,\ldots,n\}$. Denote $A_\pi$ the matrix with entries $A_{\pi(i), \pi(j)}$. Then
\be
\mathcal A = \{A_\pi:\; \pi \in \Pi\}
\ee
for a known set ({algebraic group}) of permutations $\Pi$.\footnote{If $\Pi$ is an algebraic group, then $\pi^{-1} \in \Pi$ and $\pi\circ \rho \in \Pi$ for any $\pi, \rho \in \Pi$.
  The group of permutations $\Pi$ might have more elements than the set of adjacency matrices $\mathcal A$.  
  A uniform distribution over the group nonetheless implies a uniform distribution over the matrices $\mathcal A$, due to the group structure: 
  Let $\Pi_0 = \{\pi\in \Pi:\; A_\pi = A\}$. Then there are exactly $|\Pi_0|$ elements $\rho \in \Pi$ such that $A_{\rho\circ\pi} = A_\pi$ for any $A_\pi\in \mathcal A$, so that $|\Pi| = |\mathcal A| \cdot |\Pi_0|$.
  The set $\mathcal A$ is the orbit of $A$ under the group action of $\Pi$.}

\paragraph{Exclusion restrictions and potential outcomes for networks}
We next aim to define a notion of potential outcomes for our setting. To do so requires exclusion restrictions, generalizing the ``stable unit treatment value'' assumption typically made in the context of standard discussions of binary treatment effects.\footnote{In \autoref{sec:additionaldiscussion} we review the relationship between structural functions, exclusion restrictions, the definition of potential outcomes, as well as the identification of sample average treatment effects and randomization inference, in conventional randomized experiments.}
We first introduce exclusion restrictions abstractly, before considering specific examples.

Consider some pair of nodes $(i,j)$.
Let $d_{ij}(A^1)$ be a feature or summary statistic of the network $A^1$, such as the presence of an {indirect} tie between $i$ and $j$ in period $1$.
Let $y_{ij}(A^2)$ be another feature or summary statistic of $A^2$, such as the presence of a tie between $i$ and $j$ in period $2$.
The {general form of exclusion restrictions} in our network setting is as follows:
\be
  d_{ij}(A) = d_{ij}(B) \Rightarrow y_{ij}(f(A)) = y_{ij}(f(B)),
\ee
for all $A,B \in \mathcal A$.
In words, if two networks $A, B$ are the same in terms of the feature $d_{ij}$, then the corresponding next-period networks $f(A), f(B)$ are the same in terms of the feature $y_{ij}$.
A typical example of $y_{ij}(A)$ is whether a tie is present between $i$ and $j$ at time $2$.
One example of $d_{ij}(A)$ is whether $i$ and $j$ have an indirect tie at time $1$, another example is that $d_{ij}(A)$ equals the network degree of node $i$. We discuss these examples below.

Under such an exclusion restriction, we can define the potential outcome
\be
  Y_{ij}^d = y_{ij}(f(A)) \text{ for }d_{ij}(A) = d.
\ee
The exclusion restriction implies that this definition of $Y_{ij}^d$ is independent of the choice of $A$.
For identification, we lastly also need the {support condition} that $P(d_{ij}(A^1) = d ) > 0$.\\

\subsection{Formal assumptions}
\autoref{ass:networkmodel} collects all the assumptions made thus far.
\begin{ass}[Setup]$\;$\\
\label{ass:networkmodel}
$A^1$ and $A^2$ are adjacency matrices in $\{0,1\}^{n\times n}$ satisfying the following, for some fixed $(i,j)$:
\begin{enumerate}
  \item \textbf{Structural relationship}:  $A^2 = f(A^1)$.
  \item \textbf{Repeated observation}: Both $A^1$ and $A^2$ are observed.
  \item \textbf{Randomization}: $P(A^1 = A|f) = \frac{1}{|\mathcal A|}$ for all $A \in \mathcal A$.
  \item \textbf{Exclusion restriction}: 
  $d_{ij}(A) = d \Rightarrow y_{ij}(f(A)) = Y_{ij}^d$ for all $A \in \mathcal A$.
  \item \textbf{Support}: $P(d_{ij}(A^1) = d) > 0$.
\end{enumerate}
\end{ass}

\paragraph{Permutation of nodes}
In our empirical application, the set $\mathcal A$, from which $A^1$ is drawn uniformly at random, can be obtained by {permutation} of the indices $i$ and $j$: Since new hires are randomly assigned to project teams within offices, having them (hypothetically) trade places results in a permuted initial network that is equally likely.

We will be interested in a set of statistics $d_{ij}(A^1)$ and outcomes $y_{ij}(A^2)$ indexed by $(i,j)\in \mathcal E$, for a set of edges $\mathcal E$; this will allow us to define generalizations of the sample average treatment effect ($SATE$) to the network context. For binary $d$, we might for instance consider the estimand
  $$
      \beta_{\mathcal E} = \frac{1}{|\mathcal E|}\sum_{(i,j) \in \mathcal E} \left(Y_{ij}^1 - Y_{ij}^0\right).
  $$

The statistics $d_{ij}$ that we will consider are {equivariant} under permutations.
As an example of equivariance, consider the presence of a tie, $d_{ij}(A) = A_{ij}$. Then $d_{ij}(A_\pi) = A_{\pi(i), \pi(j)} = d_{\pi(i), \pi(j)}(A)$.
  Similarly, for the network degree $dg_i$ of node $i$ it holds that $dg_i(A_\pi) = dg_{\pi(i)}(A)$.

We furthermore require that the set of edges $\mathcal E$ is {invariant} under the permutations defining $\mathcal A$. This is necessary so that estimands such as $\beta_{\mathcal{E}}$ are defined independently of the realization of $A^1 \in \mathcal A$.

\autoref{ass:permutation} captures this structure of permutations, equivariance, and invariance.
\begin{ass}[Randomization by permutation]$\;$
  \label{ass:permutation}
\begin{enumerate}
  \item \textbf{Permutations}: 
  For any permutation $\pi$ of $\{1,\ldots,n\}$, denote $A_\pi$ the matrix with $(i,j)$th entry $A_{\pi(i), \pi(j)}$.
  Let  $\Pi$ be an algebraic group of permutations.
  The set $\mathcal A$ is given by
  $$
    \mathcal A = \{A_\pi:\; \pi \in \Pi\}.%
  $$
  \item \textbf{Invariance and equivariance}:
  For all $\pi \in \Pi$, $\mathcal E$ is invariant under $\pi$, 
  $$(i,j) \in \mathcal E \Rightarrow (\pi(i),\pi(j)) \in \mathcal E,$$
  and for all  $\pi \in \Pi$ and $(i,j)\in \mathcal E$, $d_{ij}$ is equivariant under $\pi$ :
  $$
    d_{ij}(A_\pi) = d_{\pi(i), \pi(j)}(A).
  $$
\end{enumerate}

\end{ass}

\paragraph{Permutations and the support condition}
\autoref{ass:networkmodel} requires that the initial network $A^1$ is drawn uniformly at random from a set of networks $\mathcal A$. 
\autoref{ass:permutation} specializes this by assuming that $\mathcal A$ is generated by permuting nodes of the network.
In our empirical application, these permutations are obtained by permuting all the {new hires} $\mathcal I$ {within offices} $o$.
Denote $O(i)$ the office of individual $i$.
Formally, the group of permutations is then given as follows.
\begin{ass}[Permutation within subsets of nodes]
  \label{ass:rectangular}
\bals
  \Pi = \{\pi \in \mathcal{S}_n:\; &\pi(j) = j\quad \forall j \notin \mathcal I,\\
  & O(\pi(i)) = O(i)\quad \forall i \in \mathcal I\},
\eals
where $\mathcal{S}_n$ is the \emph{symmetric group} of all permutations of the set $1,\dots,n$.
\end{ass}

Given this form of $\Pi$, we can describe the maximal set of edges $\mathcal E^{max}$ for which the support condition of \autoref{ass:networkmodel} is satisfied.
For illustration, consider the case where there is only one office $o$.
Denote the theoretical support of $D_{ij}$ by $\mathcal D$, which is independent of $i$ and $j$.
Consider the matrix $(D_{ij})_{i \in \mathcal{I}, j \notin \mathcal{I}}$ of realized treatment values.
Define the set $\mathcal J$ as the set of potential ties $j$ such that $D_{ij}$ has full support across $i\in \mathcal I$.
Put differently, $\mathcal J$ is the set of column indices such that $D$ takes all possible values in $\mathcal D$, within column $j$:
\bals
  \mathcal J &= \{j\notin \mathcal I:\; \forall d \in \mathcal D\; \exists i \in \mathcal I: d_{ij}(A^1) = d\}.
\eals
Then $\mathcal E^{max}$ is the corresponding set of pairs $(i,j)$ such that $D_{ij}$ has full support given $j$:
\bals
  \mathcal E^{max} %
  &=\mathcal I \times \mathcal J.
\eals

\subsection{Examples of network characteristics $D_{ij}$}
Thus far, the setup we described is fairly abstract and general.
We next introduce specific examples of exclusion restrictions and estimands in this general setup. 
Throughout, the outcome of interest is the presence of a tie between nodes $i$ and $j$ in period $2$. We thus set
$$
Y_{ij} = y_{ij}(A^2) = A_{ij}^2.
$$

  \paragraph{Indirect ties (triadic closure)}
  
  Indirect ties have been emphasized as a determinant of tie formation in the literature on triadic closure ~\citep{Simmel1964-nr, Coleman1988-ok}. 
  A widely-held interpretation is that indirect ties---connections via mutual contacts---increase opportunities for interaction and reduce perceived risks of engagement by facilitating trust or accountability.
  
  Consider a pair $(i,j)$ of individuals who do not have a tie in period $1$, $A^1_{ij} = 0$, such that $A_{ij} = 0$ for all $A \in \mathcal A$. 
  Suppose that the only way that the network $A^1$ impacts formation of a tie between $i$ and $j$  in period $2$ is via the presence (or absence) of an {indirect tie} between $i$ and $j$ in period $1$. Formally, define the presence of an indirect tie as
  \be
    d_{ij}(A^1) = \bs 1\left(\sum_k A^1_{ik} A^1_{kj} > 0\right).
    \label{eq:indirecttie}
  \ee
  Under the exclusion restriction of \autoref{ass:networkmodel}, we can define the potential outcome $Y_{ij}^d$, for $d=0,1$. This potential outcome captures whether a tie between $i$ and $j$ would be formed in period 2, depending on whether $i$ and $j$ did or did not have an indirect tie in period $1$.

  \paragraph{Network degree (cumulative advantage)}
  Network degree has been considered as a determinant of tie formation, because greater degree increases visibility, which in turn facilitates the formation of additional ties ~\citep{Merton1968-pc, Barabasi1999-hj, DiPrete2006-ua, DiMaggio2012-th}.
  
  Consider an individual $i$.
  Suppose that the only way that $A^1$ matters for the formation of a tie between $i$ and $j$ is via the network degrees of both $i$ and $j$, where the degree of $i$ is given by
  \be
    dg_i(A^1)  = \sum_k A^1_{ik}.
    \label{eq:degree}
  \ee
  Under the permutations considered in our application, $dg_j(A)$ is invariant for all $A \in \mathcal A$. We can therefore set
  $
   d_{ij}(A^1) = dg_i(A^1).
  $

  \paragraph{Local network density}
  Local network density can shape tie formation because higher density reduces the opportunities for new additional ties to form \citep{Burt2009-oh}.

  Consider an individual $i$, as well as the individuals $k,k'$ that they are connected to.
  We define the local network density of $i$ as the share of their ties who are in turn connected to each other:
  \be
  dn_{i}(A^1) = \frac{\sum_{k>k'} A^1_{ik} A^1_{kk'} A^1_{k'i}}{ \sum_{k>k'} A^1_{ik} A^1_{k'i}}.
  \label{eq:density}
  \ee
  Local network density might influence the probability of $i$ forming additional ties.
  The local network density of $j$ is invariant under the relevant permutations, so that we consider
  $$
  d_{ij}(A^1) = dn_{i}(A^{1}).
  $$

\subsection{Relationship to parametric structural models}
Our assumptions are stated in terms of the formation of (undirected) network ties, where tie formation is based on choices of the individuals constituting the network. There is a rich econometric literature modeling the underlying preferences of individuals over tie formation, and describing the resulting networks as the equilibrium outcome of strategic interactions; see \cite{graham2020econometric} and \cite{de2020econometric} for recent reviews. Models in this literature typically rely on parametric assumptions, and often impose stronger exogeneity conditions and exclusion restrictions than we impose here. Many models in this literature are thus special cases of our more agnostic reduced-form setup; the following example illustrates.

Suppose the following: Pre-existing ties persist over time. The surplus of forming a new tie between nodes $i$  and $j$ in period $2$, given the pre-existing network in period $1$, equals

$$
U_{ij} = d_{ij}(A^{1})\cdot\delta + \epsilon_{ij},
$$
where $d_{ij}(A^{1})$ is a vector of network characteristics, as discussed above. The coefficients $\delta$ capture endogenous network mechanisms, such as triadic closure or cumulative advantage. The term $\epsilon_{ij}$ captures any unobservable determinants of tie formation. Utility is transferable between individuals. Under these assumptions, a tie is formed if and only if the surplus it generates is positive, so that

$$
A_{ij}^{2} = \mathbf{1}(d_{ij}(A^{1})\cdot\delta + \epsilon_{ij} \geq 0).
$$
In the potential outcome notation that we introduced earlier, we correspondingly have

$$
Y_{ij}^{d} = \mathbf{1}(d\cdot\delta + \epsilon_{ij} \geq 0).
$$
Note in particular that the potential outcomes $Y_{ij}^{d}$, and the function $f$ more generally, depend on the unobservables $\epsilon_{ij}$. By conditioning on the function $f$ mapping $A^{1}$ to $A^{2}$ throughout our analysis, we are implicitly conditioning on these unobservables, rather than averaging over them.

\paragraph{Homophily and triadic closure}
Let us specialize this model further to illustrate one of the key identification issues. Suppose that $D_{ij} = d_{ij}(A^{1}) = \mathbf{1}\left(\sum_k A^1_{ik} A^1_{kj} > 0\right)$, so that tie surplus only depends on the pre-existing network via the presence of indirect ties. Suppose further that
$$
\epsilon_{ij} = \exp(-|W_{i}-W_{j}|) + \eta_{ij},
$$
where $\eta_{ij}$ is iid EV1 distributed, while $W_i$ is some time invariant unobserved individual characteristic. This implies
$$
E[Y_{ij} | W_i, W_{j}, D_{ij}] = \frac{1}{1+ \exp(-[D_{ij}\cdot\delta + \exp(-|W_{i}-W_{j}|)])}.
$$
In this model, apparent triadic closure ($Cov(Y_{ij}, D_{ij})>0$) might arise for two distinct reasons: first, there is the causal effect $\delta$ of indirect ties on tie formation. Second there is the effect of unobserved homophily, where individuals who are similar in terms of $W_i$ are more likely to form ties. Because $W_i$ is time-invariant, such unobserved homophily introduces endogeneity bias in regressions of $Y_{ij}$ on $D_{ij}$; a key contribution of our approach is to address this endogeneity problem by using random variation in the initial network structure.

\paragraph{Equilibrium selection and the design-based approach}
The structural model as described above is backward-looking: Utility and the formation of ties in the network $A^2$ only depend on the network in the previous period, $A^1$.
Alternatively, we might allow for simultaneity, for instance by assuming that
$$
U_{ij} = d_{ij}(A^{2})\cdot\delta + \epsilon_{ij},
$$
Models of this form are also consistent with our framework, but raise additional subtle issues. First, they require choosing an appropriate definition of equilibrium, such as pairwise stability. Second, they typically lead to equilibrium multiplicity. Consider for instance the example of triadic closure in the previous paragraph: For $\delta$ large enough, it might be an equilibrium that all ties in a triangle form, or none of them.
In settings such as this, the structural function $f$ as defined in \autoref{ass:networkmodel} subsumes any equilibrium selection, and our estimands condition on the selection mechanism.
This contrasts with some approaches in the structural literature (e.g. \citealt{sheng2020}) that aim to estimate bounds across possible realizations of the selection mechanism.

%% file: Sections/3_estimation.tex
\section{Identification, estimation, and inference}
\label{sec:estimation}

We now turn to the construction of estimators, statistical tests, and confidence sets.
In preparation for our main results, \autoref{lem:IPW} shows that potential outcomes (as defined in \autoref{ass:networkmodel}) can be estimated without bias using inverse probability weighting, and
\autoref{lem:equivariance} characterizes the implications of invariance and equivariance (as in \autoref{ass:permutation}) for the calculation of assignment probabilities.

\autoref{prop:ipw} then shows that inverse probability weighted regressions can be used in our network setting to estimate sample average treatment effects (SATEs), and generalizations thereof.
The assumptions of \autoref{prop:ipw} are somewhat restrictive: Identification of SATEs (averaging over the set of edges $\mathcal{E}$) requires full support of treatment $D_{ij}$ for every edge $(i,j) \in \mathcal{E}$. This is hard to satisfy for richer definitions of $D$ (e.g. continuous $D$ or $D$ based on multiple network characteristics).
Addressing this issue, \autoref{prop:withinregression} shows that local average treatment effects (LATEs) can still be identified under weaker support conditions. If the strong support condition of \autoref{prop:ipw} is satisfied, then our LATEs and SATEs coincide.

We then consider computational implementation of our estimators. By appropriately sequencing calculations and storing intermediate results, estimation (and randomization inference) can be made fast even for large social networks, as in our empirical application.

We conclude this section with a discussion of inference. \autoref{prop:randomizationinference} shows how to construct finite-sample valid tests for the null of no treatment effects, using the same permutation structure that was  leveraged for identification in \autoref{prop:ipw} and \autoref{prop:withinregression}. This result parallels the main result of \cite{randomizationpeer2024}, who discuss permutation inference for peer effects, leveraging exogenous group assignment.
\autoref{prop:normality}, drawing on \cite{tian2025stratified}, shows approximate normality of the distribution of our estimators (across permutations, over counterfactual treatment assignments and counterfactual outcomes), and of the permutation test statistics (over counterfactual treatment assignments, holding outcomes constant).
\autoref{prop:standarderrors}, specializing ideas from \cite{mukerjee2018using}, derives an estimable upper bound on the standard error of our estimators over the permutation distribution, based on the variance of estimates across offices.

\subsection{Identification and estimation}

To show our subsequent propositions, we will use the following two lemmas.
All proofs can be found in Appendix~\ref{sec:proofs}.

\begin{lemma}[Unbiased estimation of potential outcomes using inverse probability weighting]
  \label{lem:IPW}
  Denote $Y_{ij} = y_{ij}(A^2)$ and $D_{ij} = d_{ij}(A^1)$.
  Under \autoref{ass:networkmodel}, the IPW estimator
$$
  \widehat Y_{ij}^d = Y_{ij} \cdot \frac{\bs 1(D_{ij} = d)}{P(D_{ij} = d )}
$$
satisfies
$$
  E\left[\widehat Y_{ij}^d|f\right] = Y_{ij}^d.
$$
\end{lemma}

\begin{lemma}[Equivariance of randomization probabilities]
  \label{lem:equivariance}
  Suppose that item 3 of \autoref{ass:networkmodel} (randomization)  and \autoref{ass:permutation} hold.
  Denote $p_{ij}(d) = P(D_{ij} = d)$, and $P_{ij} = p_{ij}(D_{ij})$.
  Then for all $\pi \in \Pi$
  $$
  p_{\pi(i)\pi(j)}(d) = p_{ij}(d), \quad
  \text{and} 
  \quad P_{\pi(i)\pi(j)} |f \sim P_{ij} | f.$$
\end{lemma}

\autoref{lem:IPW} implies that the IPW estimator is unbiased for the potential outcome $Y^d_{ij}$.
We can generalize this result by considering arbitrary {linear combinations} of potential outcomes, and their corresponding IPW estimators. 
By linearity of expectations, unbiasedness is preserved for such linear combinations. This allows us to consider estimands such as the SATE.

\begin{proposition}[Inverse probability weighted linear regression]
  \label{prop:ipw}
  Suppose that \autoref{ass:networkmodel} holds for all $(i,j) \in \mathcal E$, and that the support condition applies for every $d\in \mathcal D\subseteq\mathbb{R}^{k}$. Define $Y^{d}_{ij}$ accordingly.
  Suppose furthermore that \autoref{ass:permutation} holds.
  As in \autoref{lem:equivariance}, denote $p_{ij}(d) = P(D_{ij} = d)$, and $P_{ij} = p_{ij}(D_{ij})$.
  Define
  \bals
  \widehat \beta &= \argmin_{b} \sum_{(i,j) \in \mathcal E} \frac{1}{P_{ij}}\left(Y_{ij} - D_{ij}  \cdot b\right)^2,\\
  \beta &=   \frac{1}{|\mathcal E|}\sum_{(i,j) \in \mathcal E} \left[\left(\sum_{d\in \mathcal D} d  \cdot d'\right)^{-1}  \cdot \left(\sum_{d\in \mathcal D} Y_{ij}^d  \cdot  d\right)\right].
  \eals 
  Then 
  $$E[\widehat \beta|f] = \beta.$$
\end{proposition}

A leading example of \autoref{prop:ipw} is the case of regression on an intercept and a binary treatment, $\mathcal{D} = \{(1,0), (1,1)\}$. In this case, the estimand $\beta$ specializes to
$$
  \beta =   \frac{1}{|\mathcal E|}\sum_{(i,j) \in \mathcal E} \left(Y_{ij}^0, Y_{ij}^1  - Y_{ij}^0\right).
$$
We get in particular that the slope $\beta_2$ is equal to the SATE, $\beta_2 = \frac{1}{|\mathcal E|}\sum_{(i,j) \in \mathcal E} (Y_{ij}^1  - Y_{ij}^0)$.

Beyond this special case, \autoref{prop:ipw} also allows for multiple regressors, interactions, and projections of structural relationships on linear approximations.
Even in these more general cases, conditional on $\mathcal E$ the definition of the estimand $\beta$ remains independent of the randomization distribution; this definition only depends on the structural relationship as reflected in the potential outcomes $Y_{ij}^d$.\footnote{The randomization distribution might however enter the definition of $\beta$ to the extent that $\mathcal E$ is determined by support requirements.}

\paragraph{Local average treatment effects}
\autoref{prop:ipw} demonstrates identification of the SATE, and appropriate generalizations thereof.
Identification is based on inverse probability weighting, which requires full support of $D_{ij}$ for all $(i,j) \in \mathcal E$. 
This support condition is restrictive and does not hold in many settings of interest. 
The support condition is likely to be violated whenever $D_{ij}$ has large range (so that it can conceptually take many values), and in particular when $D_{ij}$ is defined based on multiple features of the initial network structure.

Not all is lost, however.
Even if the support condition is not satisfied, we can identify appropriate versions of a LATE. (We use the term LATE here in the generalized sense of an average causal effect where the average is taken over a population or treatment distribution determined by the assignment mechanism.) 
In the context of our application, the identity $i$ of a new hire can be interpreted as an instrument for treatment $D_{ij}$, conditional on $j$.
\autoref{prop:withinregression} shows how LATEs can be identified, using within-regressions for each potential tie $j$, across randomly assigned new hires $i$.

\autoref{prop:withinregression} weakens the support requirement, relative to \autoref{prop:ipw}, but it requires that the permutations $\Pi$ are as in \autoref{ass:rectangular}, permuting a subset $\mathcal I$ of individuals, while leaving another subset $\mathcal J$ unaffected, and that $\mathcal E = \mathcal I \times \mathcal J$. For ease of notation we consider the case of a single office $o$; aggregation across multiple offices will be discussed below.

\begin{proposition}[Within-regression]
  \label{prop:withinregression}
  Suppose that items 1-4 of \autoref{ass:networkmodel} hold for all $(i,j) \in \mathcal E = \mathcal I \times \mathcal J$, where $\mathcal I \cap \mathcal J = \emptyset$.
  Suppose furthermore that \autoref{ass:permutation} and \autoref{ass:rectangular} hold, where $O$ takes only one value.\\
   Define the multiset of treatment values $\mathcal D_j = [D_{ij}:\;i \in \mathcal I]$ (which might have repeated entries).
  Assume that $\left(\sum_{d\in \mathcal D_{j}} d \cdot d'\right)$ is invertible for all $j \in \mathcal{J}$.
  Define lastly
\bals
\widehat \gamma_{j} &=  \argmin_{c} \sum_{i \in \mathcal I}\left(Y_{ij} - D_{ij} \cdot c\right)^2,&
\widehat \gamma &=  \frac{1}{|\mathcal{J}|}\sum_{j \in \mathcal J} \widehat{\gamma}_{j},\\
\gamma_{j} &= \frac{1}{|\mathcal{I}|}\sum_{i \in \mathcal I} \left[ \left(\sum_{d\in \mathcal D_{j}} d \cdot d'\right)^{-1} \cdot \left(\sum_{d\in \mathcal D_{j}} Y_{ij}^d \cdot d\right) \right],&
\gamma &=  \frac{1}{|\mathcal{J}|}\sum_{j \in \mathcal J} {\gamma}_{j}.
\eals
Then
$$E[\widehat \gamma|f] = \gamma.$$
If additionally $\mathcal D_{j} = \mathcal D$ for all $j$, then  $\gamma = \beta$, as defined in \autoref{prop:ipw}.
\end{proposition}

To develop further intuition for the estimands $\beta$ and $\gamma$, consider the edge-specific estimands
\bals
\beta_{ij} &= \left(\sum_{d\in \mathcal D} d  \cdot d'\right)^{-1}  \cdot \left(\sum_{d\in \mathcal D} Y_{ij}^d  \cdot  d\right), &
\gamma_{ij} &=\left(\sum_{d\in \mathcal D_{j}} d \cdot d'\right)^{-1} \cdot \left(\sum_{d\in \mathcal D_{j}} Y_{ij}^d \cdot d\right).
\eals
Then $\beta$ and $\gamma$ are given by averages over the set of edges $\mathcal E$ of the edge-specific estimands $\beta_{ij}$ and $\gamma_{ij}$, respectively.
In general, the edge specific estimands are the slopes of linear approximations to the structural relationship mapping $d$ to $Y_{ij}^d$.
If the specification of $d$ is fully saturated, this recovers the actual structural mapping. If it is not, then the the approximation implicit in $\gamma_{ij}$ depends on the multi-set $\mathcal D_j$ (i.e., the conditional support); this is the sense in which $\gamma$ recovers a \textit{local} average treatment effect.

\subsection{Computation and aggregation}
\autoref{ass:rectangular} implies that only a subset of individuals $\mathcal I$ are randomly permuted, for a given office $o$.
Consider a set of edges of the form $\mathcal E =  \mathcal I \times \mathcal J$.
Given this structure, we can efficiently implement computation of the inverse probability weighting estimator of \autoref{prop:ipw}, the within-regression estimator of \autoref{prop:withinregression}, and the corresponding permutation inference of \autoref{prop:randomizationinference} (discussed below, using the notation introduced here), avoiding unnecessary repeat calculations.

\paragraph{Matrix structure}
The data can be stored in matrices of dimension $|I| \times |J|$.
Our outcome of interest is tie formation, so that $Y_{ij} = A^2_{ij}$. $Y$ is therefore the submatrix of $A^2$ corresponding to the indices in $\mathcal I \times \mathcal J$, that is, $Y = A^2_{\mathcal I,\mathcal J}$.
Similarly, $D_{ij} = d_{ij}(A^1)$, and $D$ is an array of shape $dim(D_{ij}) \times |I| \times |J|$.

Because only individuals $i\in \mathcal I$ are randomly permuted at time $1$, and because $\mathcal I \cap \mathcal J = \emptyset$, it follows that $\pi(j) = j$ for $j \in \mathcal J$, while any permutation of the indices $i \in \mathcal I$ is allowed.
$\Pi$ thus corresponds to all possible permutations of the rows of the matrix $Y$ and the array $D$, while leaving the columns unchanged.

\paragraph{Efficient calculation of the IPW estimator}
Using the notation of \autoref{prop:ipw}, 
\bals
  p_{ij}(d) &= \tfrac1{|\mathcal I|} \sum_{i' \in \mathcal I} \bs 1(D_{i'j} = d), \text{ and }&
  P_{ij} &= p_{ij}(D_{ij}).
\eals
We can efficiently calculate the probabilities $p_{ij}(d)$, which do not depend on $i$, as averages across rows $i$ for each column $j$ of $D$.
We store the $P_{ij}$ again in a matrix $P$ of shape $|\mathcal I| \times |\mathcal J|$
Define a matrix $Z$ with entries $Z_{ij} = D_{ij} / P_{ij}$, and calculate
\bal
C &= \left(\sum_{i \in \mathcal I, j \in \mathcal J} Z_{ij} \cdot D_{ij}'\right)^{-1},&
B_{i,i'} &= C \cdot \left(\sum_{j \in \mathcal J}  Z_{ij} \cdot  Y_{i'j}\right),&
\widehat \beta &= \sum_{i\in \mathcal I} B_{i,i}.
\label{eq:B}
\eal
This yields the estimator of \autoref{prop:ipw}.
The matrix $C$ is of dimension $dim(D_{ij}) \times dim(D_{ij})$, while $B_{i,i'}$ and $\widehat \beta$ are vectors of the same dimension as $D_{ij}$.
For randomization inference (discussed below), we do not need to recalculate the terms $C$ and $B_{i,i'}$ every time.
Instead, the permuted estimator $\widehat \beta_\pi$ (corresponding to permuted $D_{ij}$ but holding outcomes $Y_{ij}$ constant) is simply given by
$
\widehat \beta_\pi = \sum_{i \in \mathcal I} B_{\pi(i),i}.
$
To describe counterfactual realizations of $\hat{\beta}$, (corresponding to permuted $D_{ij}$ and the resulting counterfactual outcomes $Y_{ij}$ implied by $f$), denote
\be
\tilde B_{i,i'} = C \cdot \left(\sum_{j \in \mathcal J}  Z_{ij} \cdot  Y_{i'j}^{D_{ij}}\right).
\label{eq:Btilde}
\ee
We again have
$
\widehat \beta = \sum_{i\in \mathcal I} \tilde{B}_{i,i}.
$

\paragraph{Efficient calculation of the within-regression estimator}
A similar approach yields efficient calculation of the within-regression estimator, and the corresponding p-value. 
Define
\bal
C_j &= \left(\sum_{d\in \mathcal D_{j}} d \cdot d'\right)^{-1},&
W_{ij} &=  C_j\cdot D_{ij},&
G_{i,i'} &=\frac{1}{\lvert \mathcal{J} \rvert } \sum_{j} W_{ij} \cdot Y_{i'j}.
\label{eq:G}
\eal
Then
\bals
\widehat{\gamma}&=\sum_{i \in \mathcal{I}} G_{i,i},
\text{ and }&
\widehat{\gamma}_{\pi} &= \sum_{i \in \mathcal{I}} G_{\pi(i),i}.
\eals
Counterfactual realizations of $\hat{\gamma}$ are again described by
\be
\tilde G_{i,i'} = \frac{1}{\lvert \mathcal{J} \rvert } \sum_{j} W_{ij} \cdot Y_{i'j}^{D_{ij}},
\label{eq:Gtilde}
\ee
with $\widehat \gamma = \sum_{i\in \mathcal I} \tilde{G}_{i,i}$

\paragraph{Aggregation}
In our application, new hires are only randomly assigned to project teams within offices $o$, but not across offices.
We can, however, simply apply the argument above separately for each office, to obtain office-level estimates $\widehat \beta_o$, and permuted estimates $\widehat \beta_{o,\pi}$; similarly for $\widehat \gamma_o$ and $\widehat \gamma_{o,\pi}$.
We then define aggregate estimators
$$
\widehat \beta =   \sum_o  \frac{m_o}{m} \cdot  \widehat \beta_o,
$$
where $m_o$ is the number of new hires in office $o$, and $m$ is the total number of new hires.
We similarly define the estimand $\beta$, and the permuted estimator $\widehat \beta_\pi$, as well as $\widehat{\gamma}$, $\gamma$, and $\widehat{\gamma}_\pi$, as weighted averages across offices.
To guarantee validity of this approach, we need to ensure that $J_o \cap I_{o'} = \emptyset$ for any offices $o,o'$, i.e., that the potential ties of one office are not the new hires of another office.
The permutations $\Pi$ considered, for the whole social network, correspond to compositions of permutations of each of the subsets $I_o$ for each of the offices $o$.

\paragraph{Subsets of individuals}
Separately, we might also be interested in the interaction of pre-determined covariates with endogenous network characteristics in determining tie formation.
We might, for example, ask how the effect of triadic closure, degree, or density varies depending on the educational background, gender, or race of new hires and potential ties.
To answer questions of this form, we can simply restrict attention to subsets of $\mathcal I$ and $\mathcal J$, based on such pre-determined covariates.

We correspondingly restrict attention to random permutations within these subsets of $\mathcal I$; such permutations form a subgroup of $\Pi$. This is valid because a uniform distribution on a group of permutations induces a uniform distribution on any subgroup.
All our arguments for identification, estimation, inference, and computational implementation thus apply for the corresponding subsets $\mathcal E$ of $\mathcal E^{max}$.

\subsection{Inference}
We conclude this section by discussing the construction of hypothesis tests, standard errors, and confidence sets.
For simplicity of notation, we consider the case where $\beta$ and $\gamma$ are real-valued; the generalization to the multi-dimensional case follows immediately by adding appropriate subscripts everywhere.

\paragraph{Testing the sharp null}
The following proposition extends the idea of {randomization inference} from conventional RCTs to the network context, for the estimators of both \autoref{prop:ipw} and \autoref{prop:withinregression}.
\begin{proposition}[Randomization inference]
  \label{prop:randomizationinference}
  Under the assumptions of \autoref{prop:ipw}, consider the sharp null hypothesis that $Y^d_{ij}$ does not depend on $d$, for all $(i,j) \in \mathcal E$.\\
  For $\pi \in \Pi$, define the permuted estimator
  $$
  \widehat \beta_\pi = \argmin_{b} \sum_{(i,j) \in \mathcal E} \frac{1}{P_{\pi(i)\pi(j)}}\left(Y_{ij} - D_{\pi(i)\pi(j)}  \cdot b\right)^2,
  $$
  and the p-value
  $$
  p^\beta = \frac{1}{|\Pi|}\sum_{\pi \in \Pi} \bs 1\left(\widehat \beta \leq \widehat \beta_\pi \right).
  $$
  Under the sharp null hypothesis, 
  $$
    P(p^\beta \leq \alpha |f) \leq \alpha.
  $$
  Under the assumptions of \autoref{prop:withinregression},  define similarly
  $$
  \widehat \gamma_\pi =   
  \frac{1}{|J|}\sum_{j \in \mathcal J}
  \argmin_{c} \sum_{i \in \mathcal I}\left(Y_{ij} - D_{\pi(i)j} \cdot c\right)^2,
  $$
  and the p-value
  $
  p^\gamma = \frac{1}{|\Pi|}\sum_{\pi \in \Pi} \bs 1\left(\widehat \gamma \leq \widehat \gamma_\pi \right).
  $
  Under the sharp null hypothesis, we again get
  $
    P(p^\gamma \leq \alpha | f) \leq \alpha.
  $
\end{proposition}

\autoref{prop:randomizationinference} allows us to implement tests of the null hypothesis that $Y^d_{ij}$ does not depend on $d$.
We can sample permutations $\pi$ uniformly from the set $\Pi$, calculate $\widehat \beta_\pi$ (or $\widehat \gamma_\pi$) for each such $\pi$, and then estimate the p-value by taking a sample average of $\bs 1\left(\widehat \beta \leq \widehat \beta_\pi \right)$ (or $\bs 1\left(\widehat \gamma \leq \widehat \gamma_\pi\right)$) over these random permutations. Comparing the resulting estimated p-value to the level $\alpha$ yields a test which controls size under the null.

The p-values $p^\beta$ and $p^\gamma$ as defined in \autoref{prop:randomizationinference} are one-sided (right-tail tests). Two-sided p-values can be computed based on these one-sided p-values as $2\cdot\min(p^\beta,\, 1 - p^\beta + 1/|\Pi|)$; similarly for $\gamma$. 

\autoref{prop:randomizationinference} considers the null hypothesis of zero treatment effects, so that $Y_{ij}^d$ does not depend on $d$.
The argument is easily generalized to any null-hypothesis which allows us to impute counterfactual $Y_{ij}^d$ based on the observed data. One class of such null hypotheses, for binary $d$, posits that treatment effects are constant and equal to some known value $\delta$.

\paragraph{Approximate normality}
\autoref{prop:randomizationinference} implies that permutation inference allows us to control size exactly, under the sharp null that initial network structure does not impact subsequent tie formation.

Under our assumptions 1-3, we can characterize the distribution of our estimators further, beyond the sharp null (and allowing for full heterogeneity of effects):
The estimators $\widehat{\beta}$ and $\widehat{\gamma}$ are approximately normally distributed, conditional on the set of nodes and on the counterfactual mapping $f$, using only the randomness coming from the permutations $\pi$.  The same holds for the permutation statistics $\widehat{\beta}_{\pi}$ and $\widehat{\gamma}_{\pi}$, under the exact null of no effects.

Proposition 4 below, which formalizes this claim, follows directly from Theorem 1 in \cite{tian2025stratified}. This theorem provides generalized Berry-Esseen bounds for stratified permutation distributions. Previous results in the literature imply normality for either a large number of strata (number of offices $N$, in our setting), or a large number of observations within strata (number of new hires $m_{o}$ in an office, in our setting). Theorem 1 of \cite{tian2025stratified} provides a unified characterization.

For the following, we consider the case where the assumptions of \autoref{prop:withinregression} hold within offices $o$, offices are independent, and $\widehat{\beta},\widehat{\gamma}$ are given by weighted averages across offices. Let $\sigma^{2}_{\beta}$ and $\sigma^{2}_{\gamma}$ denote the variance of $\widehat{\beta}$ and $\widehat{\gamma}$ (given $f$, over draws of $\pi$), and let $\sigma^{2}_{\beta_{\pi}}$ and $\sigma^{2}_{\gamma_{\pi}}$ denote the variance of the permutation statistics $\widehat{\beta}_{\pi}$ and $\widehat{\gamma}_{\pi}$ (given the observed data $Y,D$, over re-draws of $\pi$).
The matrices $\tilde{B},B,\tilde{G},G$ are defined as above, in Equations \eqref{eq:B}, \eqref{eq:Btilde}, \eqref{eq:G}, and \eqref{eq:Gtilde}; recall that $m$ denotes the total number of new hires, and $m_o$ the number of new hires in an office.

\begin{proposition}[Approximate normality]
\label{prop:normality}
Under the assumptions of \autoref{prop:withinregression}, holding separately for each office $o$, consider the estimators $\widehat{\beta}$ and $\widehat{\gamma}$, and the permutation statistics $\widehat{\beta}_{\pi}$ and $\widehat{\gamma}_{\pi}$.
Denote $\bar \beta = E[\widehat{\beta}_{\pi}|Y,D]$ and $\bar \gamma = E[\widehat{\gamma}_{\pi}|Y,D]$.
Then there exists a universal constant $C$ such that
$$
\begin{aligned}
\sup_{t\in \mathbb{R}} \left| P(\widehat{\beta}\leq \beta + \sigma_{\beta} \cdot t | f) - \Phi(t) \right| 
& \leq \frac{C}{\sigma_{{\beta}}^{3}}\cdot \sum_{o} \frac{m_{o}^{2}}{m^{3}} \sum_{i,j \in \mathcal{I}_{o}} |\tilde{B}_{ij}|^{3}\\
\sup_{t\in \mathbb{R}} \left| P(\widehat{\beta}_{\pi}\leq \bar \beta + \sigma_{\beta_{\pi}} \cdot t|Y,D) - \Phi(t) \right| 
& \leq \frac{C}{\sigma_{{\beta}_{\pi}}^{3}}\cdot \sum_{o} \frac{m_{o}^{2}}{m^{3}} \sum_{i,j \in \mathcal{I}_{o}} |B_{ij}|^{3}\\
\sup_{t\in \mathbb{R}} \left| P(\widehat{\gamma}\leq \gamma + \sigma_{\gamma} \cdot t| f) - \Phi(t) \right| 
& \leq \frac{C}{\sigma_{{\gamma}}^{3}}\cdot \sum_{o} \frac{m_{o}^{2}}{m^{3}} \sum_{i,j \in \mathcal{I}_{o}} |\tilde{G}_{ij}|^{3}\\
\sup_{t\in \mathbb{R}} \left| P(\widehat{\gamma}_{\pi}\leq \bar \gamma + \sigma_{\gamma_{\pi}} \cdot t|Y,D) - \Phi(t) \right| 
& \leq \frac{C}{\sigma_{{\gamma}_{\pi}}^{3}}\cdot \sum_{o} \frac{m_{o}^{2}}{m^{3}} \sum_{i,j \in \mathcal{I}_{o}} |G_{ij}|^{3}\\
\end{aligned}
$$
\end{proposition}

The bounds in \autoref{prop:normality} are exact finite sample bounds. It is, however, instructive to consider the asymptotic behavior of these bounds under typical sampling assumptions: For appropriately bounded higher moments, the right hand side of these bounds is of order $\frac{1}{\sqrt{ m }}$, for both (1) the regime where offices are sampled independently from a super-population, and (2) the regime where the number of offices $N$ is fixed, but the number of new hires $m_{o}$ within offices diverges (cf. Corollary 4 in \citealt{tian2025stratified}).

Note again that $\sigma^2_\beta$ and $\sigma^2_\gamma$ are variances over draws of the permutation $\pi$ from $\Pi$, holding the potential outcomes $f$ fixed---not variances over hypothetical draws of $f$. This is precisely the sense in which the analysis is design-based: all randomness is attributable to the randomization of $A^1$. 
The variances $\sigma^2_{\beta_\pi}$ and $\sigma^2_{\gamma_\pi}$ similarly hold realized outcomes fixed, while re-randomizing treatment.

\paragraph{Standard errors}
Approximate normality suggests that we might construct conventional confidence intervals for $\beta$ and $\gamma$, based on estimators of the standard errors $\sigma_{\beta}$ and $\sigma_{\gamma}$. Unfortunately, these standard errors are \emph{not} identified in general (though $\sigma^{2}_{\beta_{\pi}}$ and $\sigma^{2}_{\gamma_{\pi}}$ are directly observed from the permutation distribution). Just as for conventional randomized controlled trials, these standard errors depend on the distribution (heterogeneity) of treatment effects $Y_{ij}^{1}-Y_{ij}^{0}$, cf. \cite{neyman1923application}, \cite{abadie2020sampling}, which is unknown. In particular, $\sigma_{\beta} \neq \sigma_{\beta_{\pi}}$, and the relative magnitude of these two standard deviations is undetermined, similarly for $\gamma$; cf. \cite{ding2017paradox} . That said, it is possible to estimate upper bounds on the standard errors of $\widehat{\beta}$ and $\widehat{\gamma}$. A general strategy for obtaining such upper bounds is described in \cite{mukerjee2018using}. The following proposition shows that the variance across offices gives one such upper bound. This bound is sharp if and only if treatment effects do not vary across offices.

\begin{proposition}[Conservative standard errors]
\label{prop:standarderrors}
Under the assumptions of \autoref{prop:withinregression}, holding separately for each office $o$, where $N$ is the number of offices, define
$$
\begin{aligned}
\widehat{V}_{\beta} &= \tfrac{N}{N-1}\sum_{o}  \left(\tfrac{m_{o}}{m}\widehat{\beta}_{o} - \tfrac{\widehat{\beta}}{N}\right)^{2},&
\widehat{V}_{\gamma} &= \tfrac{N}{N-1}\sum_{o}  \left(\tfrac{m_{o}}{m}\widehat{\gamma}_{o} - \tfrac{\widehat{\gamma}}{N}\right)^{2}.
\end{aligned}
$$
Then
$$
\begin{aligned}
E\left[ \widehat{V}_{\beta} |f \right]&\geq \sigma^{2}_{\beta},&
E\left[ \widehat{V}_{\gamma} |f \right]&\geq \sigma^{2}_{\gamma}.
\end{aligned}
$$
\end{proposition}

%% file: Sections/4_background.tex
\section{Empirical application: Background and data}
\label{sec:background}

\paragraph{Setting: ConsultCo}
Our empirical analysis is based on proprietary data on employees' intraorganizational networks from ConsultCo, a global professional services firm with over 50,000 employees across multiple cities and departments.
ConsultCo provides clients with expertise in areas such as management, strategy, finance, information technology, and human resources (HR) to help them improve their overall effectiveness and profitability~\citep{OMahoney2013-oy}.

The majority of ConsultCo's entry-level employees are hired through on-campus recruiting, resulting in cohorts of junior employees with relatively homogeneous educational backgrounds and profiles. 
Once hired, employees typically work on multiple project teams of varying size and duration, and collaborate to conduct research and analyses, develop solutions, and deliver recommendations to ConsultCo's clients~\citep{Morkes2023-cz}.

These project teams form the basis of employees' intraorganizational networks. We say that a tie is present between two employees if they are both working on the same project in a given month. 
Project team composition evolves continuously; employees join and leave projects throughout their duration.
These ties significantly impact the firm’s work product and employees’ career advancement, making tie formation critical, salient, and highly dynamic in this firm.

\paragraph{Team assignment}
At ConsultCo, initial assignments to project teams are centrally determined, while later collaborations are chosen by employees themselves.

An HR manager assigns new hires to their first project teams.
This initial assignment is random within offices. 
To confirm this, we conducted 29 informational interviews with both HR managers and employees. 
To provide additional support for random assignment, we furthermore report the results of a series of placebo tests in \autoref{sec:empirical} below.

After the initial assignment of new hires to project teams, subsequent assignment is based on an internal labor market. Junior employees are eager to be placed on high-priority or client-facing projects to prove their value to the firm, and senior colleagues have flexibility and discretion in who they ``recruit'' or ``select'' to work on their projects.
Subsequent project team assignment therefore reflects a bilateral choice between junior employees and senior colleagues. 

Our informational interviews with HR managers confirmed the shift from centralized initial project team assignment to network-based staffing. Interviewees highlighted that employees are explicitly told that their relationships with project teammates will shape their future opportunities to join new project teams.

\paragraph{Employment and staffing records}
We source data from seven years of administrative employment and staffing records from ConsultCo, for the years 2014 to 2020.
This allows us to study the networks of new hires entering the organization between 2015 and 2019.
The administrative employment data includes employees’ hiring dates, job titles, geographic locations, and initial departmental assignments, as well as self-reported race and ethnicity, gender, and education history. 
ConsultCo's staffing records provide data on the full set of project teams that employees report working on each month.
Following guidance from the company, we exclude project teams with more than 60 employees because they represent distinct administrative functions rather than employees working on a single project.
The composition of project teams varies over time, such that different sets of employees can work on a given project at different points in time.
Using these staffing records, we observe the project teams that employees work on each month. We use this information to construct monthly intraorganizational networks.

\paragraph{Offices, network ties, and treatment variables}
Employees of ConsultCo work in offices $o$.
For the purposes of our analysis, we define offices based on geographic location, department, and fiscal year.
Each fiscal year, a cohort of new hires starts working at an office, and gets assigned to multiple project teams.\footnote{Conditional on the definition of an office $o$, the estimates $\widehat{\beta}^o$ and $\widehat{\gamma}^o$ can be interpreted as discussed in \autoref{sec:estimation}.
Different offices $o$, however, can correspond to different fiscal years. The average effects of the form $\widehat \beta = \frac{1}{\sum_o  |\mathcal I^o|} \cdot \sum_o  |\mathcal I^o| \cdot  \widehat \beta^o$ therefore need to be understood as averages of SATEs across both organizational sub-units of ConsultCo and across multiple years.}

We define the network of an initial coworker $k$ of a new hire $i$ as the set of all colleagues that $k$ worked with in the year prior to $i$ starting at the company; cf. \autoref{tab:network}.
This timeframe aligns with prior research on workplace networks~\citep{Battilana2012-tf, McGinn2013-sc, Hansen2005-ic} and network measures used in the General Social Survey~\citep{Marsden2019-bm}.
It also captures the reality of project-based knowledge work: most projects last less than a year, so a one-year window includes both active and recently concluded collaborations while smoothing over temporary leaves (e.g., medical, parental).

We will consider several specifications of the treatment variable $D_{ij}$, based on the initial network $A^1$. \autoref{tab:treatment_definitions} summarizes these definitions.

\begin{table}[t]
\caption{Definition of network ties based on common projects}
\label{tab:network}
\vspace{12pt}
\centering
\begin{tabular}{ll}
\toprule
Variable & Definition \\
\midrule
$A^1_{ik}$ & New hire $i$ and coworker $k$ had a project in common \\
& at any time in the first three months of employment of $i$. \\
\addlinespace
$A^1_{kj}$ & Employees $k$ and $j$ (not new hires) had a project in common\\
& at any time in the 12 months prior to the new hires' collaboration with $j$. \\
\addlinespace
$A^2_{ij}$ & New hire $i$ and coworker $j$ had a project in common \\
& at any time in months 4 through 12 of $i$'s employment. \\
\bottomrule
\end{tabular}
\vspace{20pt}
\caption{Definition of treatment variables}
\label{tab:treatment_definitions}
\vspace{12pt}
\centering
\makebox[\textwidth][c]{%
\begin{tabular}{ll}
\toprule
{Variable} & {Definition} \\
\midrule
\multicolumn{2}{l}{\textsc{Continuous}} \\
\quad N indirect ties & Number of indirect ties between $i$ and $j$ (cf. \autoref{eq:indirecttie}). \\
\quad Degree & Network degree of $i$ (cf. \autoref{eq:degree}). \\
\quad Local density & Local network density (cf. \autoref{eq:density}). Scaled in percentage points. \\
\addlinespace
\multicolumn{2}{l}{\textsc{Binary}} \\
\quad Indirect tie & Indicator for whether an indirect tie was present (\textit{N indirect ties} $>0$). \\
\quad High degree & \textit{Degree} greater than the sample median of $75$. \\
\quad High density & \textit{Local density} greater than the sample median of $63$. \\
\bottomrule
\end{tabular}
}
\end{table}

\paragraph{Analysis sample}
From 2015–2020 (prior to the COVID-19 pandemic), 19,832 full-time junior professionals held the same job title in team-based departments in ConsultCo’s U.S. offices. 
We exclude 13,300 employees with prior experience at ConsultCo, who are likely to have pre-existing networks.
We further exclude 490 employees because of missing race or ethnicity data (needed for placebo tests), lack of identifying variation in coworker connections (i.e., all new hires in the same office work on the same initial project teams), or tenure shorter than one year.
This results in a final analysis sample of 6,042 full-time inexperienced new hires $i$ who joined the firm between 2015 and 2019 and worked for at least one year.
These new hires worked in 979 different offices $o$.
There are 130,686,467 pairs of new hires $i$ and prior employees $j$ in our data. 
We discuss the characteristics of this analysis sample below.

The estimators of \autoref{prop:ipw} and \autoref{prop:withinregression} average over a set $\mathcal E$  of pairs of employees $i,j$.
To define our analysis samples $\mathcal E$, we start by considering the full set of new hires $i$, and their offices $o$. From this we then construct the following subsamples. 

\textit{Full sample ($\mathcal{E}^{max}$):} This includes all pairs $(i,j)$ where $i$ is a new hire and $j$ is a prior employee, such that at least one new hire in the same office as $i$ has an indirect tie to $j$, and at least one does not. This ensures the support condition of \autoref{ass:networkmodel} is satisfied for the \textit{Indirect tie} treatment. 

\textit{Estimation samples:} For each model specification considered below, we then restrict further to the subset of $\mathcal{E}^{max}$ for which the within-office variation in $D_{ij}$ required by \autoref{prop:withinregression} is present.

%% file: Sections/5_empirical.tex
\section{Empirical results}
\label{sec:empirical}

We are now ready to take our estimators to the data, to estimate and test different theories of endogenous network formation.\footnote{The code for our empirical analysis can be found at \url{https://github.com/maxkasy/causal_inference_network_formation}.}

\input{Figures_Tables/descriptives}
\paragraph{Sample characteristics}
Our full analysis sample has 6,042 new hires $i$ in 979 different offices $o$, and we estimate the impact of initial network characteristics on tie formation for a set of 130,686,467 pairs (potential ties) $(i,j)$.
For each model specification, we consider a subset of this sample of pairs $(i,j)$ such that the support requirements of \autoref{prop:withinregression} are satisfied: We need sufficient variation of treatment $D_{i'j}$ across new hires $i'$ who are in the same office as $i$, given $j$.

\autoref{tab:sample} shows some descriptive statistics of our full estimation sample $\mathcal E^{max}$.
Slightly less than half of the new hires $i$ in the sample are female, around 5\% are Black, and about 11\% hold a degree from a top 20 university.

Among the pairs $(i,j)$ in the full estimation sample, around 0.43\% form a connection (tie) within one year, so that $Y_{ij}=1$, in our notation.
This low number is not surprising, given the large number of potential collaborators per our sample construction, relative to the number of collaborations that an employee could possibly maintain.
This average probability of tie formation of 0.43\% should be kept in mind as a reference point for assessing treatment effects below.\footnote{Throughout this section all effects are expressed as percentage-point changes in the probability of tie formation, i.e., as absolute rather than relative changes.}

The average number of \textit{Indirect ties} among the pairs $(i,j)$ is around $1.8$. The distribution of indirect ties has a long right tail, as reflected in the standard deviation of $4.54$.
The average degree (number of initial ties, or collaborators) of new hires $i$ equals $88$, again with a large dispersion (standard deviation of $62.5$).
And local networks are quite dense: On average, 65\% of pairs among the initial collaborators $k$ of a new hire $i$ are connected between themselves; this contrasts starkly with the tie formation probability of 0.43\% for pairs $(i,j)$ in the sample.
Our binarized treatment variables, by construction, have means of about $.5$ (for \textit{High degree} and \textit{High density});  \textit{Indirect tie} has a mean of $.39$.

\paragraph{Placebo tests}
Causal identification in our setting requires that new hires are randomly assigned to teams within offices.
To test the validity of this assumption, we consider a series of placebo tests. We replace the endogenous outcome $Y_{ij}$ (tie formation in period 2), in the models to be estimated below, by pre-determined, exogenous characteristics, including gender, race, and whether new hires had a degree from a top 20 university.\footnote{The variable \textit{Edu top20} is missing for a subset of new hires because employees are not required to report it.}
Validity of our approach implies that we should estimate an effect of around $0$ for these placebo regressions, and that the corresponding p-values should be random draws from the uniform distribution on $[0,1]$.

\autoref{tab:placeboestimates} reports these estimates and p-values, for each of the three pre-determined characteristics, and for each of the three binarized treatment variables $D_{ij}$; each row of the table corresponds to one specification.
As can be seen from the p-values shown in square brackets, none of these coefficients are significantly different from $0$, with the possible exception of the regression of \textit{Female} on \textit{High density}. 
To account for multiple hypothesis testing, we might apply a Bonferroni correction. The upper and lower significance cutoffs after such a correction, for a two-sided test at 5\% level across 9 regressions,  would be $.025 / 9 = .0027$ and $1 - .025/9 = .9972$. Applying these cutoffs, we would not reject the joint null that the effect of all of the  $D_{ij}$ on all of the placebo outcomes equals $0$ -- including the regression of \textit{Female} on \textit{High density}.

Complementing \autoref{tab:placeboestimates}, \autoref{fig:permutationplacebo} in \autoref{sec:additionalplots} shows plots of the permutation distribution of placebo estimates (the shaded density), and the point estimates for our sample (shown as vertical lines). In each case, the point estimates squarely fall within the range of permutated estimates.
Notably, these plots also visually confirm the claim of \autoref{prop:normality} (which provided Berry-Esseen bounds for deviations from normality): In each case the empirical distribution of permutation statistics appears normal and centered at $0$.

\paragraph{Estimates for binary treatment definitions}
Our first set of estimates, shown in \autoref{tab:effectestimates} are based on binary definitions of $D_{ij}$, and combinations thereof.
Each row in \autoref{tab:effectestimates}, as in \autoref{tab:placeboestimates} above and \autoref{tab:continuousestimates} below, corresponds to a different model specification.

\begin{figure}[p]
\caption{Confidence intervals, binary regressors}
\label{fig:estimates}
\vspace{10pt}
\includegraphics[width = \textwidth]{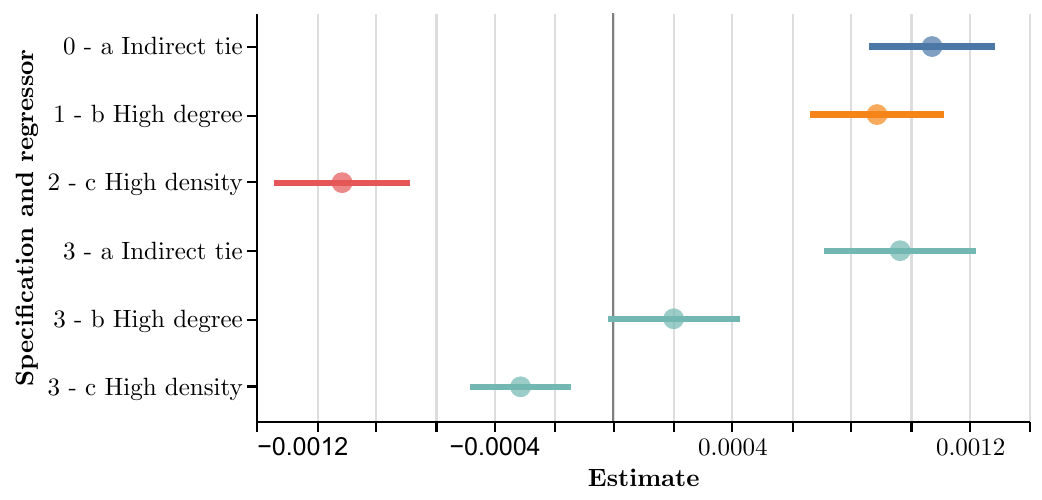}\\
\vspace{10pt}
\caption{Confidence intervals, continuous regressors}
\label{fig:continuousestimates}
\vspace{10pt}
\includegraphics[width = \textwidth]{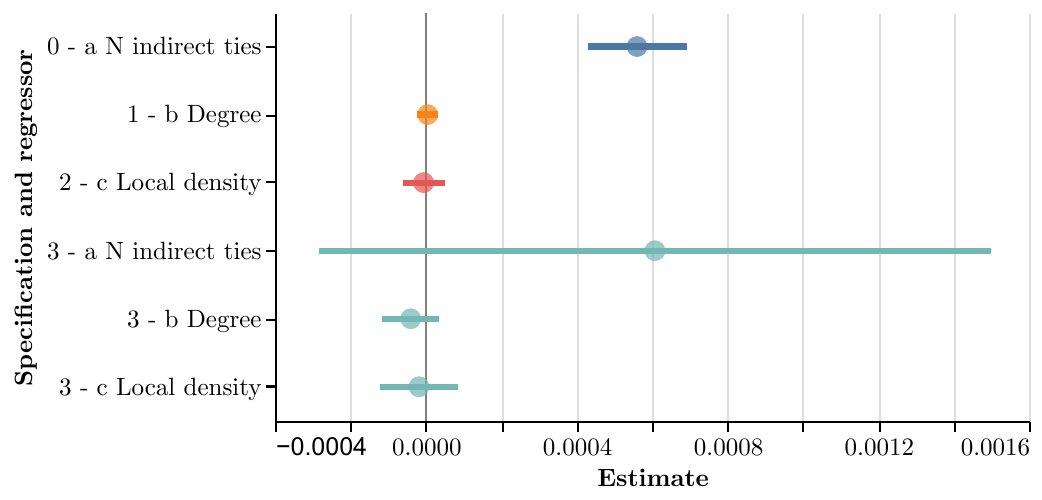}
\footnotesize
\flushleft
\textit{Notes:} These figures displays effect estimates and 95\% confidence intervals based on the standard errors shown in \autoref{tab:effectestimates} and \autoref{tab:continuousestimates}, in line with \autoref{prop:normality} and \autoref{prop:standarderrors}.

The first three estimates in each figure are based on specifications with a single regressor; the remaining three estimates are based on the joint regressions.
\end{figure}

Coefficients are shown with conservative standard errors in round brackets, and \textit{one-sided} p-values for the sharp null of no effects in square brackets, so that a p-value close to $0$ implies a significant positive effect, while a p-value close to $1$ corresponds to a significant negative effect.
\autoref{fig:estimates} and \autoref{fig:continuousestimates} show the corresponding confidence intervals.
\autoref{fig:permutationinference} and \autoref{fig:permutationinferencecontinuous} in \autoref{sec:additionalplots}  again display the full permutation distribution and point estimates.

The first three specifications in \autoref{tab:effectestimates} consider one treatment variable at a time.
In each case, the estimated coefficient is highly significant (far outside the range of the permutation distribution), and quantitatively important.
Having an \textit{Indirect tie} increases the probability of tie formation from a baseline of 0.26\% by almost 0.11\%, that is, by a factor of more than 1.4.
Taken on its own, having a \textit{High degree} similarly increases the probability of tie formation by 0.09\%, while \textit{High density} among initial collaborators appears to decrease the probability by 0.09\%.

When we consider a specification which includes all three binary treatment variables, these values change somewhat: \textit{Indirect} tie continues to have a large effect of 0.1\%. The coefficients for \textit{High degree} and for \textit{High density}, however, are considerably reduced, to values of 0.02\% and -0.03\%, respectively.

It should be noted that the estimation sample changes somewhat across different model specifications. Identification using within-regressions, as in \autoref{prop:withinregression}, requires that the matrix of $D_{ij}$ (with rows corresponding to new hires $i$ within an office $o$, and columns corresponding to treatment components) has full rank. This condition is more likely to be violated in smaller offices with fewer new hires. It also becomes more stringent as the dimension of $D_{ij}$ increases.

\paragraph{Estimates for continuous definitions of treatment}
\autoref{tab:continuousestimates} next shows the parallel set of estimates for the continuous (not discretized) definitions of the treatment variables.

As was the case for the discretized version, the number of \textit{indirect ties} has a strong and highly significant effect on tie formation.
One additional indirect tie increases the probability of a tie by 0.056\%.
Multiplied by the standard deviation of the number of indirect ties of 4.53, we get an effect of .25\% of a one standard deviation increase.

Both \textit{Degree} and \textit{Local density}, by contrast, appear to have no effect at all in this continuous linear specification. The point estimates are essentially zero, and the p-values are very far from significance.

These conclusions remain unchanged once we consider the joint model including all three continuous regressors. The effect of \textit{indirect ties} increases slightly, to 0.06\%, while \textit{degree} and \textit{local density} remain far from significant and quantitatively unimportant.

\begin{table}[p]
\centering 
\caption{Placebo tests}
\label{tab:placeboestimates}
\vspace{10pt}
\small
\begin{adjustbox}{minipage=1.15\textwidth,center}
\renewcommand{\arraystretch}{1.5}
\input{Figures_Tables/placebo_estimates}
\end{adjustbox}
\flushleft
\textit{Notes:} This table shows estimates similar to those of \autoref{tab:effectestimates}, where the endogenous outcome of tie formation has been replaced by exogenous, pre-determined characteristics. This provides placebo tests of the assumption of random initial team-assignment of new hires. P-values between .025 and .975 indicate that we do not reject the null of random assignment at conventional levels.\\[20pt]
\normalsize
\end{table}

\begin{table}[p]
\centering
\caption{Effect estimates, binary regressors}
\label{tab:effectestimates}
\vspace{10pt}
\small
\begin{adjustbox}{minipage=1.15\textwidth,center}
\renewcommand{\arraystretch}{1.5}
\input{Figures_Tables/effect_estimates}
\end{adjustbox}
\footnotesize
\flushleft
\textit{Notes:} The rows of this table shows estimates for different models of network formation, using the estimator described in \autoref{prop:withinregression}. 
Conservative standard errors, as in \autoref{prop:standarderrors}, are shown in round brackets.
P-values using permutation inference, as in \autoref{prop:randomizationinference}, are shown in square brackets.\\[10pt]
\normalsize
\centering
\caption{Effect estimates, continuous regressors}
\label{tab:continuousestimates}
\vspace{10pt}
\small
\begin{adjustbox}{minipage=1.15\textwidth,center}
\renewcommand{\arraystretch}{1.5}
\input{Figures_Tables/continuous_estimates}
\end{adjustbox}
\footnotesize
\flushleft
\textit{Notes:} This table shows estimates using the continuous counterparts of the binary treatment variables in  \autoref{tab:effectestimates}.\\[20pt]
\end{table}

%% file: Figures_Tables/descriptives.tex
\begin{table}[t]
\caption{Characteristics of analysis sample} 
\label{tab:sample}
\vspace{12pt}
\centering
\begin{tabular}{lrr}
    \toprule
    Variable & Mean & Std dev \\ 
    \midrule\addlinespace[2.5pt]
    Tie formed & 0.004 & 0.065 \\
    \addlinespace
    Indirect tie & 0.385 & 0.487 \\
    High degree & 0.507 & 0.500 \\
    High density & 0.501 & 0.500 \\
    \addlinespace
    N indirect ties & 1.573 & 4.539 \\
    Degree & 88.120 & 62.487 \\
    Local density & 65.627 & 19.352 \\
    \addlinespace
    Female & 0.461 & 0.499 \\
    Black & 0.048 & 0.214 \\
    Edu top20 & 0.113 & 0.317 \\
    \bottomrule
\end{tabular}
\footnotesize
\flushleft
\textit{Notes:} This table shows means and standard deviations for our main analysis sample.\\[20pt]
\end{table}

%% file: Figures_Tables/placebo_estimates.tex
\begin{tabular*}{\linewidth}{lrrr p{2cm}p{2cm}p{2cm}p{2cm}}
\toprule
Outcome & New hires & Offices & Edges & Intercept & Indirect tie & High degree & High density \\ 
\midrule\addlinespace[2.5pt]
Female & 6,042 & 979 & 113,535,332 & 0.4517 & 0.01155\newline (0.00726) \newline [0.066] &  &  \\
Black & 6,042 & 979 & 113,535,332 & 0.0505 & -0.00276\newline (0.00285) \newline [0.797] &  &  \\
Edu top20 & 4,958 & 856 & 89,586,721 & 0.1130 & 0.00576\newline (0.00485) \newline [0.126] &  &  \\
Female & 4,414 & 490 & 105,968,417 & 0.4460 &  & 0.03076\newline (0.01930) \newline [0.046] &  \\
Black & 4,414 & 490 & 105,968,417 & 0.0554 &  & -0.00584\newline (0.00710) \newline [0.774] &  \\
Edu top20 & 3,589 & 421 & 87,098,920 & 0.1124 &  & -0.00647\newline (0.01158) \newline [0.685] &  \\
Female & 5,161 & 654 & 118,682,052 & 0.4743 &  &  & -0.04057\newline (0.01526) \newline [0.997] \\
Black & 5,161 & 654 & 118,682,052 & 0.0555 &  &  & -0.00203\newline (0.00747) \newline [0.618] \\
Edu top20 & 4,201 & 571 & 97,247,994 & 0.1131 &  &  & -0.00405\newline (0.00917) \newline [0.676] \\
\bottomrule
\end{tabular*}

%% file: Figures_Tables/effect_estimates.tex
\begin{tabular*}{\linewidth}{lrrr p{2cm}p{2cm}p{2cm}p{2cm}}
\toprule
Outcome & New hires & Offices & Edges & Intercept & Indirect tie & High degree & High density \\ 
\midrule\addlinespace[2.5pt]
Tie formed & 6,042 & 979 & 130,686,467 & 0.0026 & 0.00107\newline (0.00011) \newline [0.000] &  &  \\
Tie formed & 4,414 & 490 & 105,968,417 & 0.0038 &  & 0.00089\newline (0.00011) \newline [0.000] &  \\
Tie formed & 5,161 & 654 & 118,682,052 & 0.0051 &  &  & -0.00091\newline (0.00012) \newline [1.000] \\
Tie formed & 3,601 & 247 & 85,078,571 & 0.0023 & 0.00097\newline (0.00013) \newline [0.000] & 0.00020\newline (0.00011) \newline [0.029] & -0.00031\newline (0.00009) \newline [0.999] \\
\bottomrule
\end{tabular*}

%% file: Figures_Tables/continuous_estimates.tex
\begin{tabular*}{\linewidth}{lrrr p{2cm}p{2cm}p{2cm}p{2cm}}
\toprule
Outcome & New hires & Offices & Edges & Intercept & N\newline indirect ties & Degree & Local\newline density \\ 
\midrule\addlinespace[2.5pt]
Tie formed & 6,042 & 979 & 127,738,642 & 0.0040 & 0.00056\newline (0.00007) \newline [0.000] &  &  \\
Tie formed & 6,038 & 977 & 130,644,513 & 0.0034 &  & 0.00000\newline (0.00001) \newline [0.386] &  \\
Tie formed & 6,036 & 976 & 130,665,151 & 0.0059 &  &  & -0.00001\newline (0.00003) \newline [0.562] \\
Tie formed & 4,768 & 440 & 113,343,057 & 0.0065 & 0.00061\newline (0.00045) \newline [0.056] & -0.00004\newline (0.00004) \newline [0.791] & -0.00002\newline (0.00005) \newline [0.617] \\
\bottomrule
\end{tabular*}